\documentclass[twocolumn]{article}
\usepackage[T2A,T1]{fontenc}
\usepackage[utf8]{inputenc}
\usepackage{multirow}
\usepackage{amsmath}
\usepackage{graphicx}
\usepackage{bm}
\usepackage{biblatex}
\usepackage{authblk}
\usepackage{caption}
\usepackage{subcaption}
\usepackage{float}
\usepackage[autostyle]{csquotes}
\usepackage[russian, english]{babel}

\author[1]{Benjamin Steel}
\author[2]{Sara Parker}
\author[3]{Derek Ruths}
\affil[1,3]{School of Computer Science, McGill University}
\affil[2]{Department of Political Science, McGill University}

{
    \makeatletter
    \renewcommand\AB@affilsepx{: \protect\Affilfont}
    \makeatother

    \makeatletter
    \renewcommand\AB@affilsepx{, \protect\Affilfont}
    \makeatother

    \affil[1]{benjamin.steel@mail.mcgill.ca}
    \affil[2]{sara.parker@mail.mcgill.ca}
    \affil[3]{derek.ruths@mcgill.ca}
}

\title{The Invasion of Ukraine Viewed through TikTok: A Dataset}
\date{}

\begin{document}

\maketitle

\begin{abstract}
    We present a dataset of video descriptions, comments, and user statistics, from the social media platform TikTok, centred around the invasion of Ukraine in 2022, an event that launched TikTok into the geopolitical arena. User activity on the platform around the invasion exposed myriad political behaviours and dynamics previously unexplored on this platform. To this end, we provide a mass-scale language and interaction dataset for further research into these processes. In this paper we conduct an initial investigation of language and social interaction dynamics, alongside an evaluation of bot detection on the platform. We have open-sourced the dataset and the library used to collect it to the public.
\end{abstract}

\section{Introduction}

TikTok, a short-form video-sharing platform with over one billion active users \cite{stokel2022tiktok}, is currently one of the largest social media platforms in the world. Despite being one of the most recent additions to the social media ecosystem, it has become an entrenched part of many people's informational and social lives. For instance, over a quarter of people below age 25 in the United States consider TikTok their primary news source \cite{matsa2022americans}. Not surprisingly, then, the platform is widely regarded to be shaping how people view major events - including the evolving Russian invasion of Ukraine. Some major media organizations have dubbed the ongoing conflict ``the first TikTok war'' \cite{dang2022tiktok,frenkel2022tiktok,paul2022tiktok,chayka2022watching, tiffany2022myth}.

Despite the appearance of a role in shaping how the war is seen around the world, little is concretely known about the nature and extent of this influence. General usage patterns and statistics (like those cited earlier) suggest that many young people in the West are building their understanding of the war from TikTok videos. Moreover, the app has also emerged as a force within the Russian media ecosystem: after Russia criminalized criticism of the military, TikTok prevented Russian users from uploading new content and seeing new content from outside Russia, thus shaping what information Russians receive and disperse in relation to the war \cite{oremus2022tiktok}. TikTok evidently has substantial power over how major world events like the invasion of Ukraine are perceived and discussed \cite{lorenz2022tiktok}, but this influence is currently poorly understood.

Conflicts in the real world precipitate conflicts between communities in online spaces. These encounters in online spaces, however, do not have to be mirrors for the hostility in the physical world. They can be places to achieve shared understanding of the problem, map the contours of underlying cultural conflict, and bring non-military populations that are living on either side of these conflicts into a position of understanding. Effectively achieving public good through understanding. TikTok is one such place where this opportunity exists, but also where we know very little about how social dynamics take place. There's urgency in understanding how communities and individuals use this medium.

Given this, we seek to investigate, and enable others to investigate, the relationship between TikTok and the social communications and interactions surrounding the invasion of Ukraine, as well as its underlying mechanics. Quantitative research as a field has generally focused on Twitter and Facebook so far \cite{haq2022twitter, chen2022tweets, la2023retrieving, pierri2022propaganda, munch2022twitter, pfeffer2023just}, and there is a severe gap in our understanding of other social media platforms. This risks not only failing to represent those who don't use Twitter or Facebook, but also leaving our hard-won and much-needed insights into the promotion of healthier social spaces specific to one or a few platforms. This is made especially pressing given the rapid movement of social media platforms from discrete follower-friends dynamics to hyper-algorithmic relationship agnostic recommendation systems. In this world, search functions are more fuzzy and user networks are more ephemeral or absent, and this requires a re-thinking of how we collect data, and how we can create methods to improve these ecosystems. This dataset focuses around a period of dramatic political and social upheaval. As our analysis shows, this exposes a variety of social dynamics that address the broader questions surrounding the role of TikTok-like platforms in society.

To create the dataset, we identified hashtags and keywords explicitly related to the conflict to collect a core set of videos (or \textquote{TikToks}). We then compiled comments associated with a subset of those videos. In total we collected approximately 9.5 thousand videos related to the invasion of Ukraine, and analysed 4.4 million comments from those videos, coming from approximately 2.6 million users.

With the dataset in hand, we conduct a preliminary analysis of social dynamics in the dataset by observing two successive levels of social structure \cite{chang2014understanding}. At the macro-level, we look at word use and the languages used on the platform; and at the meso-level, we expose semantic clusters reflecting real-world perspectives. We show that the data contains evidence of rich political and social processes at each stage. Social dynamics can be made or shaped by automated means - which are important to characterize in themselves. For this reason, in our third analysis we consider the extent to which existing bot detection methods can be used on TikTok, and in doing so, find that strong markers of bots on Twitter are not relevant on TikTok. 

To pull this dataset, we developed a TikTok web scraper, PyTok, that allows us to download a variety of data from the platform. We have open sourced PyTok on Github at \url{https://github.com/networkdynamics/pytok}. We have released the data required to recreate the dataset (in order to allow user withdrawal from the dataset) here: \url{https://doi.org/10.5281/zenodo.7926959}. We also release the scripts used to prepare this analysis here: \url{https://github.com/networkdynamics/polar-seeds}.

\section{Background}

TikTok datasets have previously been constructed via specific TikTok challenges, where a single hashtag is used to identify relevant videos \cite{surabhi2022tiktok, basch2021global}, randomly crawled \cite{jiaxiang2020building}, and purely sound based collection \cite{sachs2021tiktok} (TikTok allows searching videos by the sound used in the video). Work has started to understand political dynamics, including misinformation and political expression, on the platform (e.g., \cite{medina2020dancing,sachs2021tiktok,basch2021global}). The closest work we have found to that done here is an analysis of a smaller political event in Tulsa, Oklahoma, where videos were collected via hashtag and sound based collection \cite{bandy2020tulsaflop}. In some events on TikTok, a sound can be used across videos to signal a shared participation in a challenge or event, and they can additionally be used to search TikTok. However, in our exploration, we find no evidence of any particular shared sounds in coverage of the invasion of Ukraine on TikTok. In general, there are limited attempts at casting a broad approach to TikTok video collection over a multi-faceted and long time-scale.

We want to continue the expansion of understanding of TikTok, in the context of a major geopolitical event. Given the centrality of TikTok to the invasion of Ukraine \cite{oremus2022tiktok}, and the continuing work to understand how major geopolitical events play out on social media, data from this platform around the invasion is critical for further understanding.

Datasets on the invasion of Ukraine have been released from other platforms, such as Twitter and Facebook (e.g., \cite{haq2022twitter, chen2022tweets, la2023retrieving, pierri2022propaganda, munch2022twitter}), alongside larger scale datasets that will contain some elements of the event \cite{pfeffer2023just}. However, these platforms have been studied much more than TikTok, and expanding our understanding of this growing platform is critical.

Some work has been done on improving methods for representatively sampling from social platforms \cite{llewellyn2017distinguishing, kim2018evaluating, hino2019representing}, but to the best of our knowledge, none cover TikTok --- only Twitter. The difference between the structure and search methods of Twitter and TikTok means that how we sample will need to be re-evaluated in this new context, though ideally transferring what we can.

\section{Method}

In this section, we explain how we built the dataset. This involved four steps: (1) understanding functionally how we can collect data from TikTok, (2) designing a methodology for how to collect a broad collection of data, (3) filtering the data, and finally (4) evaluating the representativeness of this data. There is still much work to be done in future to fine-tune this process, but this is a step along that path.

\subsection{Collection Process}

We required a method that both searched TikTok content and pulled the results (whether videos, comments, or users) returned. TikTok does not yet have a widely available research API, and an initial search of social media scraping libraries revealed several available libraries, with none completely fitting our needs. Specifically, some libraries are exclusively browser automation based, which results in slow collection times. Others are entirely API-based, which is vulnerable to TikTok API changes and CAPTCHAs \cite{freelon2018computational}. 

We therefore opted to develop a library that strikes a balance between both approaches: initial requests use a web browser automation library that can fetch browser cookies without in-depth knowledge of the TikTok API, followed by API requests once the relevant cookies are stored for improved performance. We started the library \footnote{https://github.com/networkdynamics/pytok} as a fork of the TikTok-Api developed by David Teather \footnote{https://github.com/davidteather/TikTok-Api}. Our library can collect videos from hashtags, general searches, and a user's history, as well as comments from videos. Video downloading has not yet been included as a feature. 

\subsection{Collection Methodology}

\begin{table*}[h]
    \centering
    \begin{tabular}{c|c}
        Search Term & Meaning \\
         \hline
        \foreignlanguage{russian}{володимирзеленський} & Ukrainian for \textquote{Volodymr Zelensky} \\
        \foreignlanguage{russian}{славаукраїні} & Ukrainian for: \textquote{Glory to Ukraine} \\
        \foreignlanguage{russian}{россия} & Ukrainian for \textquote{Russia} \\
        \foreignlanguage{russian}{війнавукраїні} & Ukrainian for \textquote{War in Ukraine} \\
        \foreignlanguage{russian}{війна} & Ukrainian for \textquote{War} \\
        \foreignlanguage{russian}{нівійні} & Ukrainian for \textquote{No War} \\
        \foreignlanguage{russian}{нетвойне} & Russian for \textquote{No War} \\
        \foreignlanguage{russian}{зеленский} & Russian for \textquote{Zelensky} \\
        \foreignlanguage{russian}{зеленський} & Ukrainian for \textquote{Zelensky} \\
        \foreignlanguage{russian}{путинхуйло} & Russian for \textquote{Fuck Putin} \\
        \foreignlanguage{russian}{путінхуйло} & Ukrainian for \textquote{Fuck Putin} \\
    \end{tabular}
    \caption{Translations of search terms used for finding videos related to the invasion. Note the generally pro-Ukraine sentiment, due to these hashtags being both more popular and selected due to common co-occurrence with seed set hashtags.}
    \label{tab:translations}
\end{table*}

Previous work on Twitter has compared the various search collection methods of keywords, selected users, or a continuous streaming (combined with additional filtering) API, and found that the stream and user-based methods produced a more focused set of posts that are more relevant to the searched topic \cite{llewellyn2017distinguishing}. However, there is no equivalent of the streaming API for TikTok (other than the recommendation stream), and we argue that the public figure-based structure of Twitter versus the hyper-algorithmic structure of TikTok means that it would be difficult to find accounts central to a specific topic or event. Methods exist for sampling from the entire set of users on other platforms, but they are specific to a sequential user ID process as previously used by Twitter \cite{hino2019representing}. TikTok uses a timestamp and hashing-based user ID allocation process, which does not allow inference of user ID from a timestamp \cite{benson2020tinkering}. Therefore, we opt for a keyword-based search method.

Using our library, we collected all videos appearing under TikTok's search functionality (under this URL: \url{https://www.tiktok.com/tag/{search-term}}) for a range of hashtags. However, hashtag searches seem to be limited to approximately the 1000 most viewed videos. However, hashtag searches are also limited to people using a specific hashtag, and also only display the top viewed videos. 

Therefore, to expand the video set, we also used the general search functionality which allows more videos to be found for a search term, for example using the URL: \url{https://www.tiktok.com/search/video?q={search-term}}. This search comes at the cost of including more unrelated videos due to the fuzzy nature of the black-box algorithm based search functionality, and additionally videos appeared to be ordered in no available parameter based order. In some cases where a search term produced many unrelated videos, searching with a hashtag in the general search functionality improved the specificity of the search.

We used the following search terms in our video collection: \\
\texttt{standwithukraine, russia, nato, putin, moscow, zelenskyy, stopwar, stopthewar, ukrainewar, ww3, \foreignlanguage{russian}{володимирзеленський, славаукраїні, путінхуйло, россия, війнавукраїні, зеленський, нівійні, війна, нетвойне, зеленский, путинхуйло}, \#denazification, \#specialmilitaryoperation, \#africansinukraine, \#putinspeech, \#whatshappeninginukraine}. \\ 

We used a \textquote{seed-and-snowball} approach to build the list of search terms, we did an initial qualitative search of TikTok, to find the most common hashtags used in popular videos related to the war, and we used this set as the seed set of hashtags (the first ten in the list above). We collected videos tagged with these terms. We then examined the ranked co-occurring hashtags in the collected video descriptions to expand our search terms. Although we started with only English-language hashtags in the seed set, we realised that Ukrainian and Russian hashtags had the highest co-occurrences with our seed set of hashtags, so we added them as the subsequent 10 search terms in the list. These search terms are not all strictly associated with the invasion of Ukraine, but in our initial explorations we found that, in general, the majority of content collected using the terms was pertinent. See Table \ref{tab:translations} for translations of the non-English search terms.

Due to the limited search functionality of TikTok, only the most popular videos are provided as results for a search. So in order to find less popular videos, one needs to use less popular search terms. We were able to find five terms related to the invasion that had a small total view count, which we added to our search terms to get a set of videos with a range of view counts. Note these last five search terms are written with a hashtag, which is needed to increase the specificity of TikTok's fuzzy search engine for these less popular terms.

Due to the fuzzy search functionality of TikTok, some of the videos returned by the search did not contain an exact search match and in some cases were completely unrelated. For example, in the most extreme case we observed, results for \textquote{denazification} mostly returned results for \textquote{derealization}, a topic unrelated to Ukraine. In another instance, while searching for \textquote{nato} with this search method, TikTok returned videos with the keyword \textquote{naruto}, a popular anime.

To be clear, hashtag-based and general keyword searches are less-than-ideal ways of collecting data. However, these are the only available methods for searching and pulling large volumes of data from TikTok. The only alternative is the main page stream, dictated by the opaque TikTok algorithm, which is therefore even less suitable as an unbiased data source (though, admittedly keyword searches are also subject to TikTok's unknown notions of relevance and priority). Critically, TikTok does not provide a method for historical searches, with keywords being the only search filter method, ordered exclusively by popularity or black-box algorithm. This means that if we want topical datasets from TikTok, we need to obtain them as the event occurs, or else less popular videos may become available, but unfindable \cite{hino2019representing}. TikTok has released a public research API that may become widely available, but there is no indication that this will feature more advanced search methods.

With these limitations in mind, we ran this video collection process in July 2022, which has a number of implications for the final characteristics of the data. As we will see in the rest of the paper, some videos are now no longer available, there are temporal peaks in the data due to recency bias in search, and due to the limitations on historical search in TikTok, some videos will be no longer findable via the current search methods.

We repeated the collection process in April 2023, using just the first 10 search hashtags (as we found this produced videos that were more reliably related to the invasion of Ukraine, and had more videos available). 

All this points to the need for concerted future work on developing best practices for collecting data on TikTok and platforms with similar design and means of access. If we want to understand the connection between the function of TikTok and its content, and work towards improving social platforms for everyone, we will need more, solid, data.

\subsection{Filtering}

Due to the noise of the various search functionalities, and hashtag abuse, there were some irrelevant videos in the raw dataset. In this next step, we removed this irrelevant content from the data to produce the final dataset, and ensure it contains only videos related to the invasion of Ukraine. We define ``related to the invasion'' as any video or post containing or discussing one of the following items:

\begin{itemize}
    \item Depicting or discussing actual combat.
    \item Support for either side's war efforts (support or protests against a side), including propagandistic content (national anthems, footage of Ukraine in an explicit context of showing what there is to save and defending it against destruction).
    \item Any mention of Putin or Zelensky during the invasion.
    \item Coverage of critical political and military leaders engaging with the invasion (world leaders or politicians in Ukraine, Russia, US, France, UK etc.)(and their speculated objectives).
    \item Videos about direct social or economic outcomes of the war (refugees, destruction, embargoes, potential drafting, Nazis on either side, sanctions etc.)
    \item Speculation about the war (an invasion of Poland).
    \item Videos about the militaries of countries involved in the war posted during the invasion (as implied propaganda). 
\end{itemize}

Using this definition, we took samples of 100 videos at a time from the dataset, opened each video in TikTok, and manually labelled them as related or not. We found that of the videos that were still available (i.e. not deleted), 63\% were related to the invasion, with the rest being content from inside Ukraine or Russia. 29\% were no longer available. That such a high percentage of the content collected was not available only 6 months after collection signals how ephemeral much (but not all) content is on TikTok, a topic beyond the scope of this study, though certainly an important area for future work.

We trialled a few filtering strategies to find a more related subset of the videos, and after coding a few more sets of videos, fine-tuned a RoBERTa model on our coded data, using the video descriptions as input to the language model to classify the videos into being related to the invasion or not. We applied this classifier, alongside filtering to only videos from 2022 and beyond. For this filtered set, we found that of those still available, 93\% were related to the war. The unrelated videos frequently used hashtags relating to the war as a passive sign of support for some cause, or were a video from or about Ukraine, which considering we are restricting to only videos from 2022, means they are at least semi-related, but outside our definition. Of this filtered set, 30\% were no longer available, showing the high attrition rate of TikTok.

The resulting sample is highly uneven in time, with many more TikToks from the days surrounding the start of the invasion (see Figure \ref{fig:video_and_comment_counts}). However, we opted not to sample further based on any time-based method, as these have been shown to not further improve the sample's representativeness \cite{kim2018evaluating}.

For the second set of videos collected in 2023, we found that the classifier had degraded in quality for both false positives and true negatives. So we re-coded a sample from this new set, re-trained the classifier, and filtered the second set of videos, with a sample of this set producing 91\% related videos.

We combined these two filtered sets of videos and de-duplicated them. This combined set was used for the rest of our study.

With the initial video set fetched, we also collected the comments for each of these videos (limiting to 1000 comments per video to ensure reasonable collection duration), to provide additional text data for analysis.

\subsection{Evaluation}

Because TikTok data collection methods are still in their infancy, our methodology represents an important contribution to the computational study of TikTok. In particular, our methodology is flexible and allows researchers who are interested in studying other subjects or phenomena on TikTok to curate it to their needs. However, the issues we encountered with TikTok's limited search functionalities highlight how challenging unbiased sampling is with these constraints.

Understanding how to sample from a data source correctly involves understanding the general distributions of the platform. We are starting to understand the fundamental distributions of Twitter \cite{pfeffer2023just}, and therefore can determine whether a Twitter sample is representative. However, for TikTok, we will need to understand the general distributions via dataset accumulation. This means that we fundamentally do not and cannot know the representativeness of this dataset compared to the total activity around the invasion of Ukraine on TikTok, and only further research will reveal this. However, this dataset is a critical step towards understanding TikTok, enabling us to say whether this and other datasets are representative or not.

\begin{figure}
    \centering
    \includegraphics[width=\columnwidth]{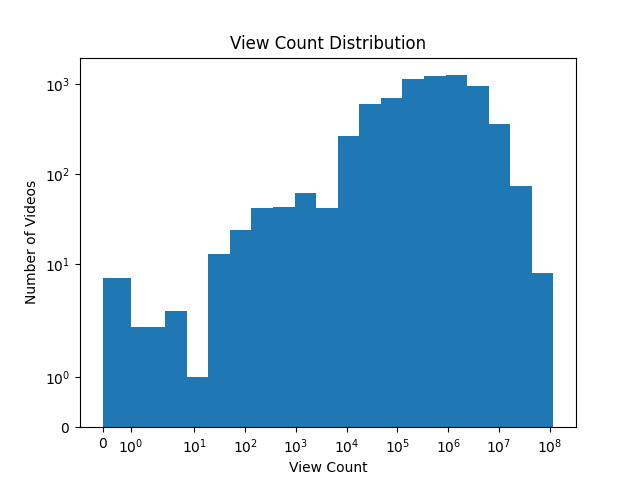}
    \caption{View count distribution of the videos in our dataset.}
    \label{fig:view_count_distribution}
\end{figure}

One measure of representativeness we consider important to improve is the distribution of view counts of videos in a dataset. The hashtag search feature, the easiest way to get data from TikTok, returns videos in descending view count order. But we would not expect that TikTok users are mostly looking at videos with the highest view counts. We can alleviate this by using the general search feature or picking less popular hashtags. See Figure \ref{fig:view_count_distribution} for the view count distribution of our collected videos.

In short, an important finding from our work in this study is the urgent need for work on best practices and methodologies for obtaining reliably representative samples of data from TikTok, since exploring how different groups use the platform must necessarily involve the analysis of more niche conversations, as well as often marginalized online communities.

\section{Dataset}

We have released the dataset, including only the data required to re-create it (to allow users to delete content from TikTok and be removed from the dataset if they wish), at the following repository: \url{https://doi.org/10.5281/zenodo.7926959}. This repository contains the unique identifiers in a CSV required to re-create the dataset, alongside scripts that will automatically pull the entire dataset. We will also include the coded sets of videos, to allow our definition of relatedness to be understood, and the relatedness classifier to be re-constructed. Once pulled using the scripts, the full dataset will be created as a collection of JSON files. There are functions available in the \textquote{PyTok} library released with this dataset to extract and assemble the most salient data from these JSON files into CSV files. Additional metadata can be found in the dataset repository.

The full dataset contains approximately 9.5 thousand videos related to the invasion of Ukraine. From the first set of videos, we were able to collect 4.4 million comments, from 2.6 million users, but this is up to the compute constraints of those downloading the dataset, and collecting more comments is certainly possible by removing max comment collection counts per video and collecting comments from more videos.

The dataset contains additional user data for users who have posted a video, including the number of users they are following, the number of followers they have, the number of videos they have, etc. Note that, unlike Twitter, TikTok does not allow others to see a user's follows. Furthermore, although a user can display the videos they have liked on their profile, this option is seldom used. This prevents methods developed for Twitter using follower or like networks from being used on TikTok data.

Of users who posted a video, the average number of videos posted was 1.7, with the max number of videos from a single user being 152. Of users who posted a comment, the average number of comments was 1.7, with a max of 832. Given that we limited our max number of comments per video to 1000, the average number of comments per video was 766. 

We plan to further expand this dataset as collection processes progress and the war continues. We will version the dataset to ensure reproducibility.

\section{Experiments}

Our objective for this study was to provide the community a dataset (and dataset collection techniques) that enables investigations into social dynamics on TikTok, around a real-world event with which the platform was deeply involved. To this end, here we offer three initial investigations to demonstrate its relevance along these lines. 

The first two investigations analyze different levels of social interaction as it pertains to the Ukraine conflict, as inspired by Chang et al. \cite{chang2014understanding}. First, we looked at the macro-level of social interaction on the platform. We wanted to understand how the relative use of words and languages changed over time, especially Russian and Ukrainian. We then zoomed in to the meso-level, from language in general to themes that emerged in posted content. Specifically, we used topic modelling to assess the extent to which conflict-relevant themes were emerging in TikTok activity.

The third investigation takes a different approach and looks at obstacles to bot detection on TikTok - which is always a central concern whenever we attempt to measure "human" behavior on social media (to be clear, bots can be considered part of the social miasma, but deserve different treatment for a variety of reasons). Bots are common on social media sites, and can skew dataset analysis without careful examination and removal. Therefore we applied a bot detection classifier to our data to try to uncover any potential bots.

\subsection{Words and Languages}

We first took a macro view of how language prominence evolved on the platform, looking at how the words used on the platform as a whole changed over time. Specifically, we wanted to measure the use of languages and words from the beginning of the invasion and onwards, to understand if there was any one changepoint at the start of the invasion, or if there were more complex, ongoing changes as communities and the platform itself responded to the unfolding crisis. The time series produced by this exposes the temporal share of language communities and the various attention cycles of the platform.

\begin{figure}
     \centering
     \begin{subfigure}[b]{\columnwidth}
         \centering
         \includegraphics[width=\columnwidth]{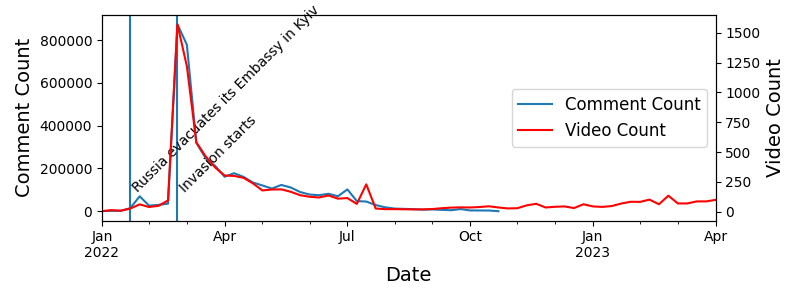}
         \caption{Count of Comments and Videos}
         \label{fig:video_and_comment_counts}
     \end{subfigure}
     \hfill
     \begin{subfigure}[b]{\columnwidth}
         \centering
         \includegraphics[width=\columnwidth]{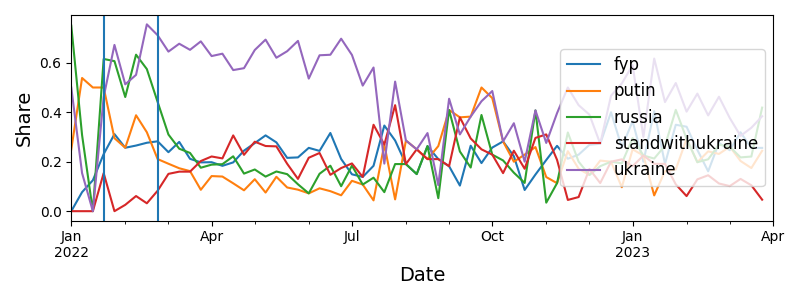}
         \caption{Top 5 Hashtag Counts}
         \label{fig:top_hashtag_counts}
     \end{subfigure}
\caption{Variety of metrics showing the change in language and languages used over the course of the invasion. (a) Number of videos and comments in the dataset over the course of 2022 and the start of 2023. Note the uptick in videos and comments around July when we did our first round of collection, due to TikTok favouring more recent content in video and comment rankings. (b) Varying usage of the top 5 hashtags in video descriptions.}
    \label{fig:videos_over_time}
\end{figure}

\begin{figure}
     \centering
     \begin{subfigure}[b]{\columnwidth}
         \centering
         \includegraphics[width=\columnwidth]{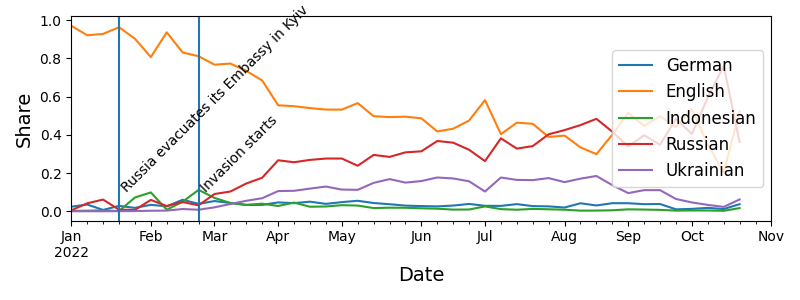}
         \caption{Top 5 Language Counts}
         \label{fig:language_counts}
     \end{subfigure}
     \hfill
     \begin{subfigure}[b]{\columnwidth}
         \centering
         \includegraphics[width=\columnwidth]{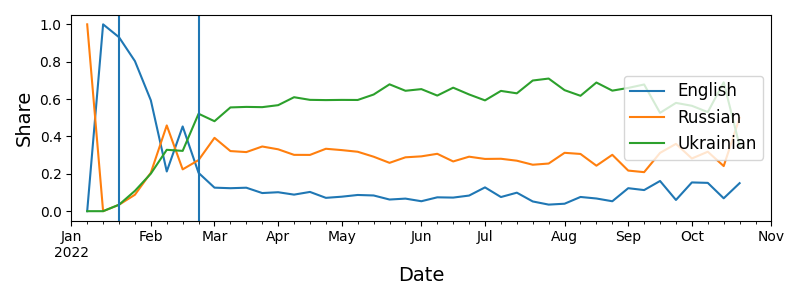}
         \caption{Language use of Users who have made at least one Comment in Ukrainian}
         \label{fig:ukrainian_language_counts}
     \end{subfigure}
     \hfill
     \begin{subfigure}[b]{\columnwidth}
         \centering
         \includegraphics[width=\columnwidth]{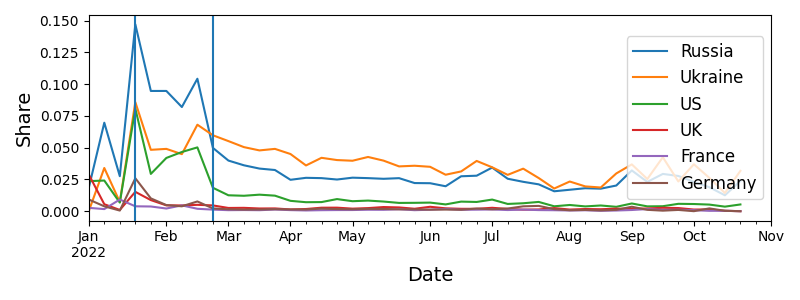}
         \caption{Country Mentions Count}
         \label{fig:country_mention_counts}
     \end{subfigure}
     \hfill
     \begin{subfigure}[b]{\columnwidth}
         \centering
         \includegraphics[width=\columnwidth]{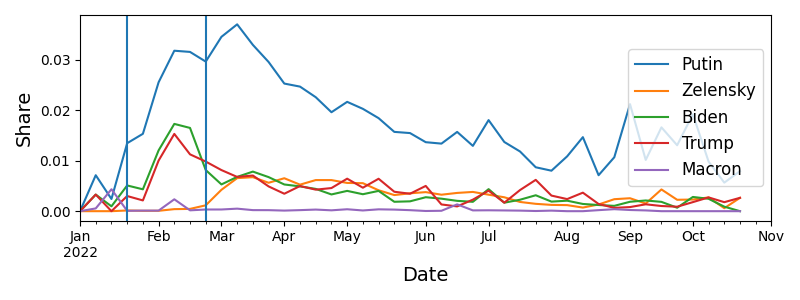}
         \caption{Leader Mentions Count}
         \label{fig:leader_mention_counts}
     \end{subfigure}
    \caption{Variety of metrics showing the change in language and languages used over the course of the invasion. (a) The top 5 languages used in comments, where language was a data attribute of comment data sent by TikTok. (b) The varying usage of Russian, Ukrainian and English by commenters who used Ukrainian at least once (as a proxy for what language Ukrainians are using on TikTok) (c) Time varying English comment mentions of various countries, where countries have multiple common names in English, we combine time data from all spellings. (d) Time varying mentions of country leaders from comment data.}
    \label{fig:comments_over_time}
\end{figure}

To do this temporal analysis we used simple keyword, hashtag, and language searches in the comments we have from the videos, with language data provided by TikTok data. Figures \ref{fig:videos_over_time} and \ref{fig:comments_over_time} feature a range of measures showing the variation in data over the time-span of the invasion. We show data on the number of videos and comments in our dataset over time, and the varying usage of the top 5 most used hashtags in video descriptions. For languages, we show the varying usage of the top 5 languages used in comments, and the varying usage of Russian, Ukrainian and English by commenters who used Ukrainian at least once (as a proxy for what language Ukrainians are using on TikTok). And for a keyword analysis, we look at the time varying mentions of countries and country leaders.

Where common keywords have multiple possible ways of saying them (USA, America for the United States of America, and Zelensky, Zelenskyy, Zelenskiy for Zelenskyy), we have searched for all of these terms, and simply added the counts to find the final search count. Note that these results are not meant to provide a comprehensive understanding of language patterns, but rather to show what possible new effects can be uncovered with this data.

\begin{figure*}[h]
    \centering
    \includegraphics[width=\textwidth]{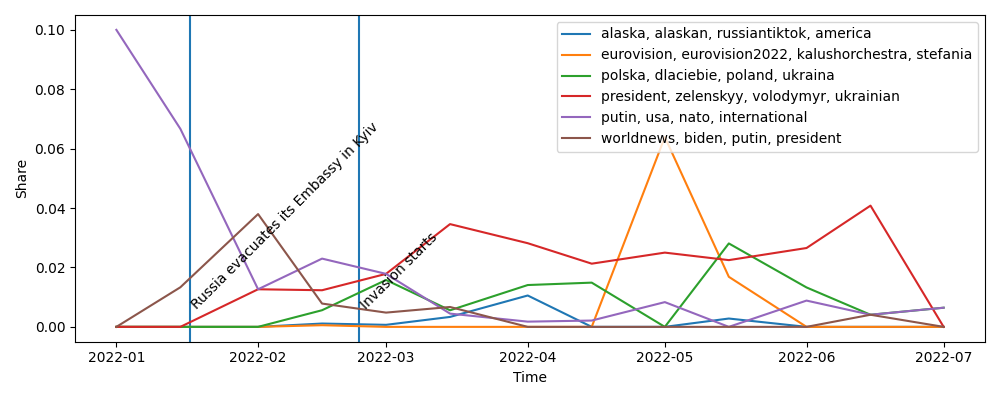}
    \caption{Frequency of a set of topics over the timeline of the invasion. The set of topics was selected for diversity of temporal and categorical nature. Topic names show the top 4 words in the topic as chosen by TF-IDF.}
    \label{fig:topic_timeline}
\end{figure*}

At this general level of social interaction, we see a number of behavioural changes over time. Notably, we see evidence of the mass movement to the Ukrainian language instead of Russian over time by users who at some point use the Ukrainian language in Figure \ref{fig:ukrainian_language_counts} \cite{afanasiev2022war}. We also see a slow change from majority English text to Russian text over the course of the invasion in Figure \ref{fig:language_counts}, indicating sustained attention from Russian speakers but a decrease in attention from English speakers. It seems likely that the medium of video allows greater mutual interaction between different language populations, engendering sophisticated temporal language dynamics that may not be seen on a text based platform.

We also examined country and leader name mentions, finding that attention was held on Putin throughout the war with only minor fluctuations, but attention on Zelenskyy diminished more rapidly in Figure \ref{fig:leader_mention_counts}. Despite heavy attention on Zelenskyy in the news, attention was not held particularly highly on him on TikTok compared to Putin. Conversely, we see sustained attention on Ukraine, but quickly diminishing attention on Russia in Figure \ref{fig:country_mention_counts}. This presents a curious juxtaposition among TikTok users: maintaining a focus on Ukraine the country and events within it while paying attention to Putin (rather than Zelenskyy). The pairing is certainly explainable, but is only one of multiple plausible pairings that could have been observed --- deserving further investigation.

In general the data clearly shows that there were a number of surprising large-scale language use changes in the run up to, at the time of, and after, the invasion, particularly of a political nature. Future research comparing these patterns to those on other platforms could further clarify the distinct characteristics of TikTok and TikTok usage.

\subsection{Topics}

As many young people increasingly use TikTok as a primary news source, it is important to understand the narratives they are consuming, and how they are shaped by the platform. Content on the platform has the power to massively shape their perception of major world events, whether it be aligned with credible news media coverage, or misinformation.

We therefore zoomed into the meso-level to find semantic clusters on the platform. For this experiment, we used topic modelling to examine video descriptions, as they provide us a view into the major themes highlighted on the platform. We used the BERTopic library \cite{grootendorst2022bertopic} and found that using a multilingual Twitter fine-tuned language model \cite{delucia2022bernice} combined with HDBSCAN based hierarchical clustering provided the clearest breakdown of topics on the platform. The clustering method used can produce a very large number of clusters depending on the dataset, so we explored a range of different maximum topics and found that 40 topics provided a healthy balance between meaningfulness and distinctness of topics.

A timeline of the frequency of a set of topics, hand-picked for diversity of temporal and categorical nature, can be found in Figure \ref{fig:topic_timeline}.

Here we can see that there is a diverse range of content reflecting various political perspectives occurring on the platform. We might expect to only see content that is aligned with perspectives in the general public media discourse, but in the topics we found, we also see popular content that likely received more attention on TikTok than in other areas. One example of this is the discussion of Alaska in the context of the invasion, a connection which saw little media attention but was clearly discussed on TikTok, indicative of discourse unique to the platform.

The frequency of topic use over time reflects a combination of temporal dynamics, from spiking at the time of invasion to increasing over the course of the war. We see discussion of Biden, and the role of the USA, and the international community spiking before the official start of the invasion, growing interest in Zelenskyy over the course of the invasion, two angles of the war that were heavily highlighted in news media, alongside the more TikTok centric discussion of Ukraine's win in Eurovision in 2022. However, other topics show steady use, including discussion of Poland's central place in the invasion. We see that TikTok features a range of discussion centred on subjects both in the media and unique to the space, indicating complex temporal meso-level processes that are causing the shifts in major topics occurring on the platform.

\subsection{Bots}

All major social media sites have a high prevalence of bots, and there is a substantial market for bots on TikTok \cite{kolomeets2021analysis}. We expect this will be an exceptionally prominent problem in the context of the Ukraine invasion, where many sides stand to gain or lose significantly from influencing public opinion. We wanted to investigate the presence of bots on TikTok, but to the best of our knowledge, no open-source bot detection tool currently exists for TikTok, with most attention paid to Twitter. To our knowledge, there are no free and open-source, out-of-the-box methods for pure content-based bot detection. We therefore applied the most recent, free, and open source library for bot detection, a state-of-the-art Twitter bot classifier focusing on generalizability \cite{ram2021birdspotter}. To operationalize the features of the classifier we used similar TikTok features such as following count, account create time, account bio. To be clear, we did not anticipate that this model would work on the out-of-distribution TikTok data, but we wanted to understand how bad the performance degradation would be.

In short, the performance was very bad. 99.8\% of accounts received a score of over 0.5, and 99.2\% received a score over 0.9, indicating that the bot detection system considers the overwhelming majority of comments on the platform as bots. Looking at the actual comments, it seems highly unlikely that nearly every comment in the dataset was generated by a bot. Examining the top features used by the gradient-boosted decision tree classifier for these classifications, we find that verified status, following count, and account age are the top contributors to this classification, alongside retweet count. As TikTok does not have a parallel retweet feature, we experimented with operationalizing this as 0, or as the like count for each comment, but this did not significantly affect the percentage of accounts with high bot probabilities, nor the top features contributing to a bot prediction. However, for the other features with direct parallels, it's clear that the predictive relationships between being a bot or not, and being verified, following count, and account age are different between TikTok and Twitter. This could indicate that there is a higher percentage of unverified accounts that are real on TikTok than on Twitter, that people tend to follow fewer accounts on TikTok (on account of the different following mechanisms), and that accounts tend to be newer on TikTok on account of it being a newer platform. 

All this underscores perhaps what is already obvious, which is that Twitter and TikTok are radically different platforms. But the exceptionally poor performance of the bot detection tooling renders clear how operationally important these differences are. As TikTok moves to take on more of a public square status \cite{guiao2021move}, many of our methods and intuitions about social media will need to be updated for this new platform, including critical misinformation classifiers, and particularly methods that rely on network data. More to the focus on this investigation, our results also reveal a need for simple baseline bot detection methods that are reliable across platforms.

\section{Conclusion}

An initial exploration of the dataset presented here indicates that there are myriad ongoing complex language dynamics on the TikTok platform related to the invasion of Ukraine. These language dynamics can be seen at macro and meso-levels of complexity, from coarse language changes to topical clusters, demonstrating that this dataset exposes a variety of salient social dynamics on TikTok. It is clear that there are variety of behaviours occurring on, and perhaps driven by, the platform, and that these behaviours can be seen in this dataset. Deriving if and how the platform drives these processes will prove fruitful for understanding how to make more participatory social spaces.

Simultaneously, issues with ensuring representative data collection and our assessment of bot detection methods on the dataset surfaced the urgent need for existing social media tools and methods to be adjusted - perhaps even re-conceptualised - to work with this new platform.

We hope that, by releasing this dataset and library, we enable researchers to work towards a more holistic understanding of the dynamics and behaviours occurring on TikTok, while also generalizing previous methods from other platforms. In particular, we hope that the presence and character of disinformation, hate speech, narrative building, filter bubble creation and polarization can be investigated using this cross-lingual dataset. And on a larger scale, we expect that the combination of this dataset with datasets from other platforms will result in stronger comparative analyses between platforms to produce a deeper comprehension of these social processes, and ultimately inform the promotion of healthier platforms.

\section{Ethical Statement}

While this dataset could allow the exposure of the influence of TikTok on its user base and improve the transparency of the platform, it also represents an analysis of personal data that TikTok users may not have fully informed consent on. Though all data analyzed in this dataset is public facing, they may not have released that data with the knowledge of all its final uses. This is why we have the released the dataset in such a way that if users want to remove the content they have on TikTok, it will no longer be able to be accessed as part of this dataset. Additionally, we have ensured that no users are identifiable through this paper, and are only identifiable through the dataset if their videos are left public.

\section{Acknowledgments}

We would like to thank Dinara Karimova for her help with translations and interpreting our results.

\printbibliography

@article{grootendorst2022bertopic,
  title={BERTopic: Neural topic modeling with a class-based TF-IDF procedure},
  author={Grootendorst, Maarten},
  journal={arXiv preprint arXiv:2203.05794},
  year={2022}
}

@article{freelon2018computational,
  title={Computational research in the post-API age},
  author={Freelon, Deen},
  journal={Political Communication},
  volume={35},
  number={4},
  pages={665--668},
  year={2018},
  publisher={Taylor \& Francis}
}

@article{stokel2022tiktok,
 author  = {Stokel-Walker, Chris},
 date    = {2022-02-21},
 title   = {TikTok Wants Longer Videos - Whether You Like It Or Not},
 journal = {Wired},
 url     = {https://www.wired.com/story/tiktok-wants-longer-videos-like-not/},
 urldate = {2022-02-21}
}

@article{dang2022tiktok,
 author   = {Dang, Sheila and Culliford, Elizabeth},
 date     = {2022-03-07},
 title    = {TikTok war: How Russia's invasion of Ukraine played to social media's youngest audience},
 journal  = {Reuters},
 url      = {https://www.reuters.com/technology/tiktok-war-how-russias-invasion-ukraine-played-social-medias-youngest-audience-2022-03-01/},
 urldate  = {2022-03-07}
}

@article{frenkel2022tiktok,
 author    = {Frenkel, Sheera},
 date      = {2022-03-05},
 title     = {TikTok is Gripped by Violence and Misinformation of Ukraine War},
 journal   = {New York Times},
 url       = {https://www.nytimes.com/2022/03/05/technology/tiktok-ukraine-misinformation.html},
 urldate   = {2022-03-05}
}

@article{benson2020tinkering,
 author  = {Benson, Ryan},
 date    = {2020-09-11},
 title   = {Tinkering with TikTok Timestamps},
 journal = {dfir.blog},
 url     = {https://dfir.blog/tinkering-with-tiktok-timestamps/},
 urldate = {2023-05-03}
}

@article{munch2022twitter,
 author     = {Munch, Felix and Kessling Philipp},
 date       = {2022-02-25},
 title      = {Twitter data around the Ukraine Invasion in February 2022},
 journal    = {OSF HOME},
 url        = {https://doi.org/10.17605/OSF.IO/RTQXN},
 urldate    = {2023-05-03}
}

@article{paul2022tiktok,
 author    = {Paul, Kari},
 date      = {2022-03-20},
 title     = {TikTok was ‘just a dancing app.’ Then the Ukraine war started},
 journal   = {The Guardian},
 url       = {https://www.theguardian.com/technology/2022/mar/19/tiktok-ukraine-russia-war-disinformation},
 urldate   = {2022-03-20}
}

@article{chayka2022watching,
 author    = {Chayka, Kyle},
 date      = {2022-03-03},
 title     = {Watching the World’s First TikTok War},
 journal   = {The New Yorker},
 url       = {https://www.newyorker.com/culture/infinite-scroll/watching-the-worlds-first-tiktok-war},
 urldate   = {2022-03-03}
}

@article{tiffany2022myth,
 author    = {Tiffany, Kaitlyn},
 date      = {2022-03-10},
 title     = {The Myth of the First TikTok War},
 journal   = {The Atlantic},
 url       = {https://www.theatlantic.com/technology/archive/2022/03/tiktok-war-ukraine-russia/627017/},
 urldate   = {2022-03-10}
}

@article{oremus2022tiktok,
 author    = {Oremus, Will},
 date      = {2022-04-13},
 title     = {TikTok created an alternate universe just for Russia},
 journal   = {The Washington Post},
 url       = {https://www.washingtonpost.com/technology/2022/04/13/tiktok-russia-censorship-propaganda-tracking-exposed/},
 urldate   = {2022-04-13}
}

@article{matsa2022americans,
 author    = {Matsa Eva, Katherina},
 date      = {2022-10-21},
 title     = {More Americans are getting news on TikTok, bucking the trend on other social media sites.},
 journal   = {Pew Research Centre},
 url       = {https://www.pewresearch.org/fact-tank/2022/10/21/more-americans-are-getting-news-on-tiktok-bucking-the-trend-on-other-social-media-sites/},
 urldate   = {2022-10-21}
}

@article{afanasiev2022war,
 author    = {Afanasiev, Ievgen},
 date      = {2022-04-24},
 title     = {The war has many Ukrainians who speak Russian abandoning the language},
 journal   = {NPR},
 url       = {https://www.npr.org/2022/04/24/1094567906/the-war-has-many-ukrainians-who-speak-russian-abandoning-the-language},
 urldate   = {2022-04-24}
}

@article{lorenz2022tiktok,
 author    = {Lorenz, Taylor},
 date      = {2022-03-11},
 title     = {The White House is briefing TikTok stars about the war in Ukraine},
 journal   = {The Washington Post},
 url       = {https://www.washingtonpost.com/technology/2022/03/11/tik-tok-ukraine-white-house/},
 urldate   = {2022-03-11}
}

@article{chang2014understanding,
  title={Understanding the paradigm shift to computational social science in the presence of big data},
  author={Chang, Ray M and Kauffman, Robert J and Kwon, YoungOk},
  journal={Decision Support Systems},
  volume={63},
  pages={67--80},
  year={2014},
  publisher={Elsevier}
}

@inproceedings{kolomeets2021analysis,
  title={Analysis of the malicious bots market},
  author={Kolomeets, Maxim and Chechulin, Andrey},
  booktitle={2021 29th conference of open innovations association (FRUCT)},
  pages={199--205},
  year={2021},
  organization={IEEE}
}

@inproceedings{ram2021birdspotter,
  title={Birdspotter: A Tool for Analyzing and Labeling Twitter Users},
  author={Ram, Rohit and Kong, Quyu and Rizoiu, Marian-Andrei},
  booktitle={Proceedings of the 14th ACM International Conference on Web Search and Data Mining},
  pages={918--921},
  year={2021}
}

@inproceedings{surabhi2022tiktok,
  title={TikTok for good: Creating a diverse emotion expression database},
  author={Surabhi, Saimourya and Shah, Bhavik and Washington, Peter and Mutlu, Onur Cezmi and Leblanc, Emilie and Mohite, Prathamesh and Husic, Arman and Kline, Aaron and Dunlap, Kaitlyn and McNealis, Maya and others},
  booktitle={Proceedings of the IEEE/CVF Conference on Computer Vision and Pattern Recognition},
  pages={2496--2506},
  year={2022}
}

@article{jiaxiang2020building,
  title={Building domain specific lexicon based on TikTok comment dataset},
  author={Jiaxiang, Hao},
  journal={arXiv preprint arXiv:2012.08773},
  year={2020}
}

@inproceedings{medina2020dancing,
  title={Dancing to the partisan beat: A first analysis of political communication on TikTok},
  author={Medina Serrano, Juan Carlos and Papakyriakopoulos, Orestis and Hegelich, Simon},
  booktitle={12th ACM conference on web science},
  pages={257--266},
  year={2020}
}

@article{sachs2021tiktok,
  title={The TikTok self: Music, signaling, and identity on social media},
  author={Sachs, Jeffrey and Wise, Rahshemah and Karell, Daniel},
  journal={SocArXiv. https://doi. org/10.31235/osf. io/2rx46},
  year={2021}
}

@article{basch2021global,
  title={A global pandemic in the time of viral memes: COVID-19 vaccine misinformation and disinformation on TikTok},
  author={Basch, Corey H and Meleo-Erwin, Zoe and Fera, Joseph and Jaime, Christie and Basch, Charles E},
  journal={Human vaccines \& immunotherapeutics},
  volume={17},
  number={8},
  pages={2373--2377},
  year={2021},
  publisher={Taylor \& Francis}
}

@article{guiao2021move,
  title={" Move carefully and discuss things": Taking back our public square},
  author={Guiao, Jordan},
  journal={AQ-Australian Quarterly},
  volume={92},
  number={4},
  pages={20--27},
  year={2021}
}

@article{haq2022twitter,
  title={Twitter dataset for 2022 russo-ukrainian crisis},
  author={Haq, Ehsan-Ul and Tyson, Gareth and Lee, Lik-Hang and Braud, Tristan and Hui, Pan},
  journal={arXiv preprint arXiv:2203.02955},
  year={2022}
}

@article{chen2022tweets,
  title={Tweets in time of conflict: A public dataset tracking the twitter discourse on the war between Ukraine and Russia},
  author={Chen, Emily and Ferrara, Emilio},
  journal={arXiv preprint arXiv:2203.07488},
  year={2022}
}

@article{la2023retrieving,
  title={Retrieving false claims on Twitter during the Russia-Ukraine conflict},
  author={La Gatta, Valerio and Wei, Chiyu and Luceri, Luca and Pierri, Francesco and Ferrara, Emilio},
  journal={arXiv preprint arXiv:2303.10121},
  year={2023}
}

@article{pierri2022propaganda,
  title={Propaganda and Misinformation on Facebook and Twitter during the Russian Invasion of Ukraine},
  author={Pierri, Francesco and Luceri, Luca and Jindal, Nikhil and Ferrara, Emilio},
  journal={arXiv preprint arXiv:2212.00419},
  year={2022}
}

@article{pfeffer2023just,
  title={Just Another Day on Twitter: A Complete 24 Hours of Twitter Data},
  author={Pfeffer, Juergen and Matter, Daniel and Jaidka, Kokil and Varol, Onur and Mashhadi, Afra and Lasser, Jana and Assenmacher, Dennis and Wu, Siqi and Yang, Diyi and Brantner, Cornelia and others},
  journal={arXiv preprint arXiv:2301.11429},
  year={2023}
}

@inproceedings{llewellyn2017distinguishing,
  title={Distinguishing the wood from the trees: Contrasting collection methods to understand bias in a longitudinal brexit twitter dataset},
  author={Llewellyn, Clare and Cram, Laura},
  booktitle={Proceedings of the International AAAI Conference on Web and Social Media},
  volume={11},
  number={1},
  pages={596--599},
  year={2017}
}

@article{hino2019representing,
  title={Representing the Twittersphere: Archiving a representative sample of Twitter data under resource constraints},
  author={Hino, Airo and Fahey, Robert A},
  journal={International journal of information management},
  volume={48},
  pages={175--184},
  year={2019},
  publisher={Elsevier}
}

@article{kim2018evaluating,
  title={Evaluating sampling methods for content analysis of Twitter data},
  author={Kim, Hwalbin and Jang, S Mo and Kim, Sei-Hill and Wan, Anan},
  journal={Social Media+ Society},
  volume={4},
  number={2},
  pages={2056305118772836},
  year={2018},
  publisher={SAGE Publications Sage UK: London, England}
}

@inproceedings{delucia2022bernice,
  title={Bernice: A multilingual pre-trained encoder for twitter},
  author={DeLucia, Alexandra and Wu, Shijie and Mueller, Aaron and Aguirre, Carlos and Resnik, Philip and Dredze, Mark},
  booktitle={Proceedings of the 2022 conference on empirical methods in natural language processing},
  pages={6191--6205},
  year={2022}
}

@article{bandy2020tulsaflop,
  title={\# tulsaflop: A case study of algorithmically-influenced collective action on tiktok},
  author={Bandy, Jack and Diakopoulos, Nicholas},
  journal={arXiv preprint arXiv:2012.07716},
  year={2020}
}

\end{document}